\documentstyle[twoside]{article}

\catcode`\@=11
\long\def\@makefntext#1{
\protect\noindent \hbox to 3.2pt {\hskip-.9pt  
$^{{\eightrm\@thefnmark}}$\hfil}#1\hfill}		

\def\@makefnmark{\hbox to 0pt{$^{\@thefnmark}$\hss}}	
	
\def\ps@myheadings{\let\@mkboth\@gobbletwo
\def\@oddhead{\hbox{}
\rightmark\hfil\eightrm\thepage}   
\def\@oddfoot{}\def\@evenhead{\eightrm\thepage\hfil
\leftmark\hbox{}}\def\@evenfoot{}
\def\sectionmark##1{}\def\subsectionmark##1{}}

\oddsidemargin=\evensidemargin
\newcounter{sectionc}\newcounter{subsectionc}\newcounter{subsubsectionc}
\renewcommand{\section}[1] {\vspace{12pt}\addtocounter{sectionc}{1} 
\setcounter{subsectionc}{0}\setcounter{subsubsectionc}{0}\noindent 
	{\tenbf\thesectionc. #1}\par\vspace{5pt}}
\renewcommand{\subsection}[1] {\vspace{12pt}\addtocounter{subsectionc}{1} 
	\setcounter{subsubsectionc}{0}\noindent 
	{\bf\thesectionc.\thesubsectionc. {\kern1pt \bfit #1}}\par\vspace{5pt}}
\renewcommand{\subsubsection}[1] {\vspace{12pt}\addtocounter{subsubsectionc}{1}
	\noindent{\tenrm\thesectionc.\thesubsectionc.\thesubsubsectionc.
	{\kern1pt \tenit #1}}\par\vspace{5pt}}
\newcommand{\nonumsection}[1] {\vspace{12pt}\noindent{\tenbf #1}
	\par\vspace{5pt}}

\newcommand{\textlineskip}{\baselineskip=13pt}
\newcommand{\smalllineskip}{\baselineskip=10pt}

\def\eightcirc{
\begin{picture}(0,0)
\put(4.4,1.8){\circle{6.5}}
\end{picture}}
\def\eightcopyright{\eightcirc\kern2.7pt\hbox{\eightrm c}} 

\newcommand{\copyrightheading}[1]
	{\vspace*{-2.5cm}\smalllineskip{\flushleft
        {\footnotesize Nuovo Cimento B 114 (January 1999) 107-112 #1}\\
       {\footnotesize Los Alamos electronic archives: gr-qc/9411053 #1}\\
        {\footnotesize $\eightcopyright$\,
        1999 by Nuovo Cimento \& H.C. Rosu
        }\\
	 }}

\def\abstracts#1#2#3{{
	\centering{\begin{minipage}{4.5in}\baselineskip=10pt\footnotesize
	\parindent=0pt #1\par 
	\parindent=15pt #2\par
	\parindent=15pt #3
	\end{minipage}}\par}} 

\renewenvironment{thebibliography}[1]
	{\frenchspacing
	 \ninerm\baselineskip=11pt
	 \begin{list}{\arabic{enumi}.}
        {\usecounter{enumi}\setlength{\parsep}{0pt}     
	 \setlength{\leftmargin 12.7pt}{\rightmargin 0pt} 
         \setlength{\itemsep}{0pt} \settowidth
	{\labelwidth}{#1.}\sloppy}}{\end{list}}

\newcounter{itemlistc}
\newcounter{romanlistc}
\newcounter{alphlistc}
\newcounter{arabiclistc}

\def\@citex[#1]#2{\if@filesw\immediate\write\@auxout
	{\string\citation{#2}}\fi
\def\@citea{}\@cite{\@for\@citeb:=#2\do
	{\@citea\def\@citea{,}\@ifundefined
	{b@\@citeb}{{\bf ?}\@warning
	{Citation `\@citeb' on page \thepage \space undefined}}
	{\csname b@\@citeb\endcsname}}}{#1}}

\newif\if@cghi
\def\cite{\@cghitrue\@ifnextchar [{\@tempswatrue
	\@citex}{\@tempswafalse\@citex[]}}
\def\citelow{\@cghifalse\@ifnextchar [{\@tempswatrue
	\@citex}{\@tempswafalse\@citex[]}}
\def\@cite#1#2{{$\null^{#1}$\if@tempswa\typeout
	{IJCGA warning: optional citation argument 
	ignored: `#2'} \fi}}

\def\@refcitex[#1]#2{\if@filesw\immediate\write\@auxout
	{\string\citation{#2}}\fi
\def\@citea{}\@refcite{\@for\@citeb:=#2\do
	{\@citea\def\@citea{, }\@ifundefined
	{b@\@citeb}{{\bf ?}\@warning
	{Citation `\@citeb' on page \thepage \space undefined}}
	\hbox{\csname b@\@citeb\endcsname}}}{#1}}

\def\@refcite#1#2{{#1\if@tempswa\typeout
        {IJCGA warning: optional citation argument
	ignored: `#2'} \fi}}

\def\refcite{\@ifnextchar[{\@tempswatrue
	\@refcitex}{\@tempswafalse\@refcitex[]}}

\def\pmb#1{\setbox0=\hbox{#1}
	\kern-.025em\copy0\kern-\wd0
	\kern.05em\copy0\kern-\wd0
	\kern-.025em\raise.0433em\box0}

\def\fnt#1#2{\footnotetext{\kern-.3em
	{$^{\mbox{\scriptsize #1}}$}{#2}}}

\def\runninghead#1#2{\pagestyle{myheadings}
\markboth{{\protect\footnotesize\it{\quad #1}}\hfill}
{\hfill{\protect\footnotesize\it{#2\quad}}}}
\font\tenrm=cmr10
\font\tenit=cmti10 
\font\tenbf=cmbx10
\font\bfit=cmbxti10 at 10pt
\font\ninerm=cmr9

\font\eightrm=cmr8

\def\qed{\hbox{${\vcenter{\vbox{			
   \hrule height 0.4pt\hbox{\vrule width 0.4pt height 6pt
   \kern5pt\vrule width 0.4pt}\hrule height 0.4pt}}}$}}

\begin{document}

\runninghead{Rosu, frequency spectra
$\ldots$} {Rosu, fundamental quantum effects
$\ldots$}


\normalsize\textlineskip
\thispagestyle{empty}
\setcounter{page}{1}

\copyrightheading{}			

\vspace*{0.88truein}

\centerline{\bf REMARKS ON THE FREQUENCY SPECTRA
     }
\centerline{\bf OF SOME FUNDAMENTAL QUANTUM EFFECTS}
\vspace*{0.035truein}
\vspace*{0.37truein}
\vspace*{10pt}
\centerline{\footnotesize H. ROSU}
\vspace*{0.015truein}
\centerline{\footnotesize\it Instituto de F\'{\i}sica,
Universidad de Guanajuato, Apdo Postal E-143, 37150 Le\'on, Gto, Mexico}
\vspace*{0.225truein}

\vspace*{0.21truein}
\abstracts{{\bf Summary}: -
Short remarks on the problem of assigning frequency spectra to Casimir,
sonoluminescence, Hawking, Unruh, and quantum optical squeezing effects are
presented.
}{}{}


\textlineskip                  
\vspace*{12pt}                 

\vspace*{1pt}\textlineskip	
\vspace*{-0.5pt}
\noindent


\noindent




\noindent

{\bf Introduction}

\noindent
Recent years witnessed great progress toward understanding
in a quantum field approach the following fundamental quantum effects:
Casimir, sonoluminescence,
Hawking, Unruh, and squeezing ones. These effects have been treated in many
ways, various authors emphasizing either peculiar aspects or similarities
between them. A large body of knowledge was accumulated over the years
in their regard but nevertheless they continue to be highly challenging
topics. Many well-known authors retain special interest in these
paradigm-like effects and their treatment is pervading a good deal of
the modern literature in theoretical physics.

\noindent
A unifying basis of all these fundamental effects may occur when approaching
them either from the standpoint of quantum vacuum energy, i.e., the energy of
zero-point fluctuations [\refcite{vac}], or as quantum Brownian
noises [\refcite{hm}].
Such treatments are very helpful for those
who would like to stress the similarity of such effects. It is by now
well settled in the literature that the vacuum energy may have in
certain conditions a thermal-like representation. This is a most
interesting fact by itself, and a glimpse to the active
developments of more precise scientific terminology for the general concept
of {\it thermality} will convince the reader on the progress achieved
since the times of Wien and Planck. Let us point out for example that at
least three kinds of reservoirs have been recognized, depending on the noise
they produce. One may speak of reservoirs with coloured noise, with
phase-dependent noise and of squeezed-vacuum reservoirs. It is true however
that the majority of the analytical treatments are based on the white-noise
assumption, which is appropriate whenever the correlation time of
the reservoir is small as compared to the dynamical time scale. This
is essentially the case in all the practical situations.

\noindent
The heat-bath features of vacuum fluctuations within the realm of the
mentioned effects show
similarities of course, but also differences, and it is the main
purpose of the present work to draw attention, at a heuristic level, on
such dichotomies that in our
opinion may prove quite useful for further progress in the understanding
of those effects.
For example, the analogy between Casimir effect
and Hawking effect has been under the focus of a number of authors.
One of the most detailed analysis of this analogy has
been provided by Nugayev [\refcite{nu}], and is based on the approach
according to which the blackbody radiation in the Hawking case is
created in the close vicinity of the event horizons.
On the other hand, Grishchuk and Sidorov [\refcite{gs}], dealing with the
squeezing of the gravitational waves, have briefly considered
the squeezing produced by a Schwarzschild black hole. They showed that
the Bogolubov-Hawking transformations can be interpreted as a
two-mode type of squeezing, a conclusion that applies to
Rindler motion as well. Thus, it appears
that there are definite connections among the various standpoints.
A good procedure for studying the interplay of the effects
is to examine the assignment of a frequency spectrum in each case because in
this way one may hope to
better characterize the differences as well as the similarities
between them. In our opinion, the frequency spectrum is an extremely
appropriate means to disentangle the dichotomy
{\it similar-different} for such effects.
Frequency spectra are the basic means to investigate stationary noises
and even nonstationary ones, though the situation is a bit delicate in the
latter case for which tomographical methods may be the best [\refcite{mm}].
As is known the
integral of the frequency spectrum is called the noise power. This is the
basic concept used both in theoretical and engineering considerations. Let us
consider the spectrum of Schottky noise, which is of
primary importance for stochastic cooling in storage
rings [\refcite{tot}]. This noise expresses the fluctuation of electrostatic
potential near the beam. Its spectrum is thus directly related to the
charge-density fluctuation in the beam which in turn
is characterized by static and dynamic form factors. But by means of the
fluctuation-dissipation theorem the dynamic form factor is related
to the dielectric response function. All of these relationships are very
useful when dealing with noises, either classical or quantum ones.
Especially in the latter case it is known that the Fourier transform of the
Kubo-Martin-Schwinger condition [\refcite{kms}] is actually the thermal
fluctuation-dissipation relation.

\noindent
Our remarks in the following refer to the Casimir
effect, sonoluminescence, Hawking radiation and Unruh radiation,
quantum optical squeezing, and also to frequency measurements of uniformly
accelerating observers, in this order, ending up with conclusions.

\bigskip

{\underline{\it 1. Frequency Spectrum for Casimir Effect}}

\noindent
Although the total zero-point energy of the electromagnetic field
contained in a cavity bounded by conducting walls is divergent, its
variation due to the displacement of the boundaries is finite and
corresponds to a weak but measurable attraction between the walls (for a
review see [\refcite{p}]).
This was first remarked by Casimir in 1948. In the case of two parallel,
conducting plates separated by a distance $d$, Casimir obtained the
 simplest expression for the Casimir energy, i.e., the interaction
 energy density per unit area as follows
 $$
 \rho _{C}(d)=-\frac{\pi ^{2}}{720}\frac{\hbar c}{d^{3}}~.
 \eqno(c.1)
 $$

\noindent
Ford [\refcite{1}] was the first to pose the problem of the frequency
spectrum in the case of Casimir effect. He defined the spectrum
by means of suited spectral weight functions distorting the
original spectrum of quantum fluctuations. In this way, such functions
are able to reveal the contribution of each frequency interval to the
 finite Casimir energy. His motivation was based on the example of the
 energy density of a massless scalar field in $S^{1}\times R$, the
  two-dimensional flat spacetime with spatial periodicity of length $L$.
  A work of Brevik and Nielsen [\refcite{bn}] on the Casimir energy of a
  piecewise string is also very close to this simple model.
  The Casimir energy density can be expressed as follows
  $$
  \rho =-\pi ^{-1}
  \int_{0}^{\infty} \frac{\omega d\omega}{e^{L\omega}-1} =
  -\frac{\pi}{6L^{2}}~.
  \eqno(c.2)
  $$
  As remarked by Ford a suggestive interpretation is that of an integral
  over a thermal spectrum with a temperature equal to $L^{-1}$.
  However, many different integrals over $\omega$ could yield the same
  result. More examples of energy-momentum tensors that can be
  written as integrals with a thermal denominator but different
  phase-space numerators are given in [\refcite{c}].
  As a toy model for the effects of the spectral weighting Ford
  used the vacuum energy density of a scalar field in $S^{1}\times R$
$$
\rho_{W}=(2L)^{-1}\sum_{n=-\infty}^{n=+\infty}\omega _{n}
W(\omega_{n})~.
\eqno(c.3)
$$
If the spectral weight function $W(\omega)$ vanishes sufficiently
rapidly for $\omega\rightarrow \infty$ then the weighted vacuum energy
density is finite. Ford found complex behavior of the spectrum,
with discontinuities and oscillations.
Also, according to Hacyan et al. [\refcite{hj}] the weight
function depends significantly on the specific experiment.

\bigskip

\underline{{\it 2. Spectrum of sonoluminescence}}

\noindent
Sonoluminescence (SL) is an intriguing phenomenon known since 1934 and
consisting in picosecond flashes
of light that are synchronously generated by the extremely nonlinear
{\em cavitation collapse} of, e.g., water bubbles which are trapped at the
velocity node of a resonant sound field in water. Recent experiments on
water SL have been performed by Hiller, Putterman and
Barber [\refcite{son}]. These authors claim that SL is radiation
of black-body type at a temperature as high as 25,000 K. There are many other
experiments and the whole topic is in a very active period.
The last scientific works of Schwinger
attributed SL to a dynamical Casimir effect [\refcite{Sch}], but recent
debate do not favor this interpretation.
The SL phenomenon
can be extrapolated to other fields of physics, for example, one may think of
a SL-type phenomenon as responsible for the gamma bursts in astrophysics.

\bigskip

\underline{{\it 3. Spectrum of Hawking Radiation and of Unruh Radiation}}

\noindent
A paper of H. Ooguri [\refcite{o}]
under the title ``Spectrum of Hawking radiation and the Huygens principle"
has as purpose to discuss a result of Takagi [\refcite{t}] who proved that
the vacuum spectrum detected by a uniformly accelerating detector in the
case of a free massless scalar field in
Minkowski spacetime (i.e., the Unruh effect), and similar spectra in de Sitter
and Schwarzschild spacetimes are
of Bose-Einstein type in even spacetime dimensions and of Fermi-Dirac type
in odd number of spacetime dimensions.
According to Ooguri this is a consequence of the fact that massless
field theories in odd dimensions do not satisfy the quantum version of
the Huygens principle, i.e., the expectation value of the commutator of
massless fields does not vanish in the timelike region.
Takagi proved that the vacuum power spectrum of a massless scalar field
in $n$ spacetime dimensions could be written in the following form
$$
F_{n}(\omega)=\frac{\pi}{\omega}
\frac{D^{M}_{n}(\omega)d_{n}(\omega)}{e^{\omega /T} -(-1)^{n}}~.
\eqno(h.1)
$$
$D^{M}_{n}(\omega)$ is the Minkowski density of states, i.e., that of massless
Minkowski quasi-particles in $n$- dimensional Minkowski spacetime which is
given by
$$
D^{M}_{n}(\omega)=\frac{2^{2-n}\pi ^{(1-n)/2}}{\Gamma ((n-1)/2)}
|\omega|^{n-2}
\eqno(h.2)
$$
and
$d_{n}(\omega)=D^{R}_{n}(\omega)/D^{M}_{n}$ if $n$ is even and the same
ratio multiplied by $\coth (\omega/2T)$ if $n$ is odd. The $D^{R}_{n}$ is
the Rindler density of states of Rindler quasiparticles. This result can
be extended to Hawking effect as well.
As emphasized by Takagi the statistics -changing phenomenon refers
to the distribution function characterizing the power spectrum of the noise
and not to the change in the basic algebra obeyed by the operators.
It is related to special relationships between the Green's functions in
succesive dimensions. Indeed, it is well known [\refcite{k}] that the
Wronskian condition on the coefficients of Bogolubov transformations ensures
in both Lorentzian and Euclidean spacetimes the preservation of
the formal commutation relations at any time. This has also to do
with methods of constructing wavefronts of wave equations in curved
spaces or spaces with special causal structures, in other words,
with mathematical aspects of Huygens' principle
[\refcite{io}]. Further progress in this field can be sought in terms of
Radon-type transforms [\refcite{rad}].

\bigskip

\underline{{\it 4. Spectrum of Squeezing}}

\noindent
Since, as was mentioned in the introduction, the Hawking effect, as well as
the Unruh effect, might be
interpreted as two-mode squeezing effects, we present the laboratory
squeezing case in view of further possible relationships [\refcite{ghb}].
For a general non-squeezing context, it is recommendable for the reader to
take a look in an old paper of Eberly and W\'odkiewicz [\refcite{ew}].

\noindent
The spectrum of squeezing in laboratory quantum optics has been discussed
in some detail by Carmichael [\refcite{car}], who first defined it as
the ratio of the homodyne spectrum
and the shot spectrum in the zero-frequency limit where the shot-spectrum is
flat. It is a photocurrent spectrum and for a direct comparison with
frequency spectra of the other sections one should pass to the photocount
regime. The spectrum of squeezing worked out by Carmichael is given by his
formula 3.26 in [\refcite{car}]
$$
S(\omega,\theta)=8\eta r (2\gamma _{1})
\int _{0} ^{\infty} d\tau cos\omega \tau
<:\Delta X_{\theta}(0)\Delta X_{\theta}(\tau):>~,
\eqno(s.1)
$$
where the quantity in the brackets denotes the normally ordered, time-ordered
correlation function for the intracavity field quadrature $X_{\theta}$, with
the phase $\theta$ controlled by the phase of the local-oscillator; the
$2\gamma _{1}$ factor gives the rate for photon escape through the mirror and
is a damping constant in the master operator equation of the cavity;
the parameter $r$ is the reflection coefficient at the beam splitter
while $\eta$ is a detection efficiency. The
formula $(s.1)$ is a quite standard one for a power spectrum with the
normally ordered treatment of the autocorrelation function.
According to Carmichael, the signature of squeezing at a certain frequency
$\omega$ is the phase-dependence of the power spectrum between a {\em negative}
minimum value and a positive maximum one at a phase in quadrature with
respect to the first one.

\noindent
We recall here that the detection of an amplitude component of a field
can be implemented by means of a homodyne detector. The procedure is as
follows. The signal beam is
combined by means of a beam splitter with an intense {\em local oscillator}
(LO) field operating at the same frequency. The combined field is then
directed to a photodetector and the amplitude component of the field is
revealed as the beating between the two input fields. To avoid noise
from intensity fluctuations of the LO, the balanced configuration is
usually chosen, by using two photodetectors with equal gains and a
50-50 (\%) beam splitter. The difference photocurrent $I_{D}$ between the two
photodetectors is not influenced by the fluctuations of the LO
intensity. On the contrary, $I_{D}$ measures the interference between
 the signal beam and the LO, the interference being constructive
at one photodetector and destructive at the second one [\refcite{d'a}].

\bigskip

\underline{
{\it 5. Frequency Measurements of Uniformly Accelerating Observers}}

\noindent
Measuring frequencies in noninertial frames is by far a non-trivial issue.
Recently, Moreau [\refcite{m}] commented on the nonlocality in the frequency
measurements of uniformly accelerating observers. The extension of
special relativity to accelerated frames is based on the standard
assumption of locality, that is the equivalence between an accelerated
observer and an instantaneously comoving inertial observer. However
the measurement of the frequency of a wave associated with a particle
performed by an accelerating observer is an example of a nonlocal observation
[\refcite{b}], complicating a lot the physical discussion.

\bigskip

{\bf Conclusion}

\noindent
The problem of assigning frequency spectra to various noises of
classical and quantum origin is of the first importance.
Here, I addressed this problem for a number of effects of fundamental character
in theoretical physics,
merely as an introduction to other people's
works. In this `noise' approach, the {\em thesis} is to consider
whatever the effects as noises and see what information comes out from
their noise spectra.

\nonumsection{Acknowledgement}
\noindent
This work was partially supported by the CONACyT Project
458100-5-25844E.


\newpage
\nonumsection{References}


\end{document}